\documentclass[aps,pre,amsmath,lengthcheck,superscriptaddress]{revtex4-2}

\usepackage{graphicx}\graphicspath{ {figures/} }
\usepackage{hyperref}
\usepackage{amssymb}
\hypersetup{colorlinks,allcolors=blue,breaklinks}

\newcommand{\be}{\begin{equation}}
\newcommand{\ee}{\end{equation}}
\newcommand{\ba}{\begin{align}}
\newcommand{\ea}{\end{align}}
\newcommand{\bi}{\begin{itemize}}
\newcommand{\ei}{\end{itemize}}

\newcommand{\bla}{bla\\bla\\bla\\bla\\bla}

\begin{document}

\title{Fluctuation theorems for quasistatic work}

\author{Pierre Naz\'e}
\email{pierre.naze@icen.ufpa.br}

\affiliation{\it Universidade Federal do Par\'a, Faculdade de F\'isica, ICEN,
Av. Augusto Correa, 1, Guam\'a, 66075-110, Bel\'em, Par\'a, Brazil}

\date{\today}

\begin{abstract}

When a thermally isolated system performs a driving process in the quasistatic regime, its variation of average energy is equal to its quasistatic work. Even though presenting this simple definition, few attempts have been made to describe such quantity from the fluctuation theorem point of view. In this work, based on Jarzynski's equality, four forms of such equality are deduced. To corroborate the result, a relation with the strong inequality $\langle W\rangle\ge \langle W_{\rm qs}\rangle$ is pursued. It is concluded in the end that any of the fluctuation theorems deduced cannot derive such a postulate. Also, no contradiction is observed if the strong inequality breaks down.

\end{abstract}

\maketitle

\section{Introduction} 

Fluctuation theorems of all types have been proved over the last decades \cite{sevick2008,esposito2009,seifert2012,shiraishi2015fluctuation,crooks1999entropy,iyoda2017fluctuation,malek2017,malek2017fluctuation,mandal2017entropy,pal2017integral,aaberg2018fully,lostaglio2018quantum,dabelow2019irreversibility,hasegawa2019fluctuation,micadei2020quantum,de2022quantum,bonancca2022fluctuation,mahault2022topological,murashita2022review,salazar2022thermodynamic}. They relate non-equilibrium quantities with equilibrium ones, showing the new face of the Second Law of Thermodynamics, expressed now as an equality. Historically speaking, Jarzynski's work~\cite{jarzynski1997} puts in evidence such relation, deducing from it the weak inequality for the average work performed in a thermally isolated system with its difference in Helmholtz's free energy between its final and initial equilibrium states, $\langle W\rangle\ge\Delta F$. 

Another important quantity for a thermally isolated system is the quasistatic work $\langle W_{\rm qs}\rangle$, obtained as the variation of average energy produced in a driving process performed in the quasistatic regime~\cite{acconcia2015,bonancca2016non,bonancca2019approaching,naze2022series,yi2021}. In particular, the Second Law of Thermodynamics for this kind of system is a postulated stronger inequality, $\langle W\rangle \ge \langle W_{\rm qs}\rangle$, which is tighter than the weak one since $\langle W_{\rm qs}\rangle\ge \Delta F$~\cite{jarzynski1997}. Understanding why such inequality holds is important for theoretical nonequilibrium statistical mechanics. However, as far as I know, only one attempt was made to prove a fluctuation theorem that justifies such strong inequality of thermally isolated systems: in Jarzynski's work~\cite{jarzynski2020}, the inequality is proven for macroscopic thermodynamic systems, using as an example of the ideal gas with a moving piston. In the end, the author states that this problem of finding a proof for the strong inequality for generic systems is still open.

In this work, I explore such a problem by proving four fluctuation theorems related to the quasistatic work. All of them are deduced based on Jarzynski's equality. However, none can derive the strong inequality.

\section{Problem and proposed solution}

Consider a thermally isolated system with a Hamiltonian $\mathcal{H}({\bf z}({\bf z_0},t),\lambda(t))$, dependent on some time-dependent external parameter $\lambda(t)=\lambda_0+g(t)\delta\lambda$, with $g(0)=0$ and $g(\tau)=1$, where $\tau$ is the switching time of the process. Here, ${\bf z}$ is the point in the phase space evolved from the initial point ${\bf z_0}$ until the instant of time $t$ according to the solution of Hamilton's equations. Initially, the system is in thermal equilibrium at a temperature $T=(k_B \beta)^{-1}$, where $k_B$ is Boltzmann's constant. When the process starts, the system is decoupled from the heat bath and adiabatically evolves in time, that is, without any source of heat. The average work performed along the process is
\be
\langle W \rangle(\tau) = \int_0^\tau \langle \partial_\lambda \mathcal{H} \rangle(t)\dot{\lambda}(t)dt,
\ee
where $\langle\cdot\rangle$ is the average concerning the non-equilibrium distribution of the system~\cite{jarzynski1997}. In particular, for quasistatic processes, where $\tau\rightarrow\infty$, we called it quasistatic work
\be
\langle W_{\rm qs} \rangle= \lim_{\tau\rightarrow\infty} \langle W \rangle(\tau). 
\ee
Looking at the strong inequality, one may guess that both quantities would obey a fluctuation theorem
\be
\langle e^{-\beta W}\rangle = e^{-\beta \langle W_{\rm qs}\rangle},
\ee
from which Jensen's inequality will furnish the desired relation. However, this fluctuation theorem does not make any sense at first sight. Indeed, if $W$ is the work performed by the thermally isolated system, then $\langle W_{\rm qs}\rangle$ must necessarily be the difference in Helmholtz's free energy between the final and initial equilibrium states $\Delta F$ of this system. This only happens when the system evolves with a heat bath around its surroundings.

However, four modifications in this fluctuation theorem to make it valid can be considered: 1) $\langle W_{\rm qs}\rangle $ is a type of $\Delta F$, but referring to another system than the original one; 2) the exponential is changed by another increasing and convex function; 3) the initial probability distribution is different than the canonical one; 4) Jarzynski's equality is considered for two averages relating the work random variable $W$ and quasistatic work random variable $W_{\rm qs}$. Thus, by using Jensen's inequality, one recovers relations where the strong inequality can be explored.

\section{First fluctuation theorem} 

I start by defining a system of interest, with a not-known yet Hamiltonian $\mathcal{G}({\bf w}({\bf w_0},t),\lambda(t))$, which has the same time-dependent external parameter of the original system. Again, ${\bf w}$ is the point in the phase space evolved from the initial point ${\bf w_0}$ until the instant of time $t$ according to the solution of Hamilton's equations of such system.

To determine the Hamiltonian $\mathcal{G}$, I consider the regime of thermal equilibrium since the demonstration of Jarzynski's equality requires only the knowledge of equilibrium states of the system~\cite{jarzynski1997}. First, the quasistatic work is an integrated function of the exact differential of the energy $E_{\rm qs}$. Therefore
\be
\langle W_{\rm qs}\rangle = \langle E_{\rm qs} \rangle_0|_{\lambda_0}^{\lambda_0+\delta\lambda},
\ee
where $\langle\cdot\rangle_0$ is the canonical ensemble of the Hamiltonian $\mathcal{H}({\bf z_0},\lambda_0)$ calculated at temperature $T$. The partition function of the system of interest is
\be
\mathcal{Z}(\lambda_0) = \int e^{-\beta \mathcal{G}({\bf w_0},\lambda_0)}d{\bf w_0}=e^{-\beta \langle E_{\rm qs} \rangle_0}.
\label{eq:partitionfunction}
\ee
I propose the Hamiltonian $\mathcal{G}$ is given by
\be
\mathcal{G}({\bf z}(t),\lambda(t))=\mathcal{H}({\bf z}(t),\lambda(t))+\langle E_{\rm qs}\rangle_0(\lambda(t))-F(\lambda(t)),
\ee
where $F$ is Helmholtz's free energy of the original system. It is easy to see that Eq.~\eqref{eq:partitionfunction} is satisfied. Observe also that the system of interest has the same solution as the original system, and the same probability distribution. This implies the fluctuation theorem~\cite{jarzynski1997}
\be
\langle e^{-\beta W'}\rangle = e^{-\beta \langle W_{\rm qs}\rangle},
\label{eq:1stft}
\ee
where 
\be
W' = \mathcal{G}({\bf z}({\bf z_0},\tau),\lambda_0+\delta\lambda)-\mathcal{G}({\bf z_0},\lambda_0).
\ee
By using Jensen's inequality, one has
\be
\langle W'\rangle \ge \langle W_{\rm qs}\rangle.
\label{eq:wadwqs}
\ee
In particular, using in Eq.~\eqref{eq:wadwqs} the inequality $\langle W_{\rm qs}\rangle\ge\Delta F$~\cite{jarzynski2020}, one obtains the following relation
\be
\langle W'\rangle \ge \Delta F.
\label{eq:waddeltaf}
\ee
Summing up the weak inequality with Eq.~\eqref{eq:wadwqs}, one has 
\be
\langle W\rangle+\langle W' \rangle\ge \langle W_{\rm qs}\rangle +\Delta F,
\ee
from which one can deduce
\be
\underbrace{\langle W'\rangle-\Delta F}_{\ge 0}\ge \langle W\rangle-\langle W_{\rm qs}\rangle \ge \underbrace{\Delta F-\langle W'\rangle}_{\le 0},
\ee
showing that this fluctuation theorem does not offer a tight bound to the difference $\langle W\rangle-\langle W_{\rm qs}\rangle$. Also, admitting this difference to be negative, there is no contradiction with Eq.~\eqref{eq:waddeltaf}.

\section{Second fluctuation theorem}

For the second fluctuation theorem for $\langle W_{\rm qs}\rangle$, I propose the following relation
\be
\langle f(-\beta (W-\langle W_{\rm qs}\rangle)) \rangle = 1,
\label{eq:2ndftwqs}
\ee
where $f(x)$ would be a globally convex and globally increasing function, possibly dependent of $\delta\lambda/\lambda_0$ and $\tau$, and with $f(0)=1$. Indeed, if it holds, the strong inequality is proved by Jensen's inequality. Let us consider that it holds
\be
\langle f(-\beta (W-\langle W_{\rm qs}\rangle)) \rangle=\langle e^{-\beta (W-\Delta F)} \rangle,
\label{eq:2ndftwqs}
\ee
Expanding in Taylor's series around $x=0$, one finds that
\be
f^{(n)}(0)=\frac{\langle (W-\Delta F)^n\rangle}{\langle (W-\langle W_{\rm qs}\rangle)^n\rangle},
\ee
which indicates that the function $f(x)$ is a perturbed exponential, since $W_{\rm qs}=\Delta F+\epsilon$, with $\epsilon\ge 0$. Observe that in this case $f(0)=1$, as required. Also, that $f(x)$ depends on $\tau$ and $\delta\lambda/\lambda_0$. To be globally convex and globally increasing, it is necessary to be locally convex and locally increasing at all points. Let us see the situation around the point $x=0$. The second derivative is
\be
f''(0)\ge 0,
\ee
so the function is locally convex around $x=0$. However, the first derivative is
\be
f'(0)=\frac{\langle W\rangle-\Delta F}{\langle W\rangle-\langle W_{\rm qs}\rangle},
\ee
which requires the previous knowledge of the sign of the difference $\langle W\rangle-\langle W_{\rm qs}\rangle$ to know if the function is locally increasing or not. Therefore, the fluctuation theorem deduced cannot derive the strong inequality. Actually, if the difference is negative, the definition of the function $f(x)$ opens the possibility to agree with such an aspect in terms of the increasing property.

\section{Third fluctuation theorem}

The third fluctuation theorem for $\langle W_{\rm qs}\rangle$ proposed is the following relation
\be
\overline{e^{-\beta (W-\langle W_{\rm qs}\rangle)}}=1,
\ee
where $\overline{\cdot}$ is an average over an initial probability distribution different than the canonical one.

First, one can always define a probability distribution by defining all its moments~\cite{feller1991introduction}. In this way, considering
\be
\overline{e^{-\beta (W-\langle W_{\rm qs}\rangle)}} = \langle e^{-\beta (W-\Delta F)} \rangle,
\ee
one expands both sides in Taylor's series around $x=0$, establishing the following definitions for the moments of the new probability distribution
\be
\overline{(X-\langle W_{\rm qs}\rangle)^n}:=\langle (X-\Delta F)^n\rangle,
\ee
for any random variable $X$ and $n\ge 0$. In particular, for $X=W$, one has
\be
\overline{W}-\langle W_{\rm qs}\rangle=\langle W\rangle-\Delta F\ge 0.
\ee
Summing up with $-\langle W_{\rm qs}\rangle$ and using the previous inequality, one arrives at
\be
\underbrace{\overline{W}-\langle W_{\rm qs}\rangle}_{\ge 0}\ge \langle W\rangle -\langle W_{\rm qs}\rangle \ge \underbrace{\Delta F-\langle W_{\rm qs}\rangle}_{\le 0}, 
\ee
showing that the strong inequality can not be proven again. Observe that assuming $\langle W\rangle-\langle W_{\rm qs}\rangle$ negative does not imply any contradiction with the inequalities $\overline{W}-\langle W_{\rm qs}\rangle$ and $\Delta F - \langle W_{\rm qs}\rangle$.

\section{Fourth fluctuation theorem}

The last fluctuation theorem is a direct consequence of Jarzynski's equality
\be
\langle e^{-\beta W}\rangle=e^{-\beta \Delta F}=\langle e^{-\beta W_{\rm qs}}\rangle.
\ee
Using Jensen's inequality, one can deduce
\be
\underbrace{-\frac{1}{\beta}\ln{\left(\frac{e^{-\beta\langle W\rangle }}{\langle e^{-\beta W}\rangle}\right)}}_{\ge 0}\ge\langle W\rangle-\langle W_{\rm qs}\rangle\ge \underbrace{-\frac{1}{\beta}\ln{\left(\frac{\langle e^{-\beta W_{\rm qs}}\rangle}{e^{-\beta\langle W_{\rm qs}\rangle }}\right)}}_{\le 0}.
\label{eq:4thft}
\ee
Again, a fluctuation theorem was found out, but the strong inequality can not be derived. Indeed, the relation above illustrates the reason behind such difficulty: for a thermally isolated system, the probability distribution of $W_{\rm qs}$ is not a Dirac delta centered at $\langle W_{\rm qs}\rangle$, which opens the possibility of the desired difference to be negative.

The second inequality of Eq.~\eqref{eq:4thft} suggests an attempt to prove the strong inequality. Considering
\be
f(\delta\lambda)=\langle W\rangle-\langle W_{\rm qs}\rangle,
\ee
and
\be
g(\delta\lambda)= -\frac{1}{\beta}\ln{\left(\frac{\langle e^{-\beta W_{\rm qs}}\rangle}{e^{-\beta\langle W_{\rm qs}\rangle }}\right)},
\ee
observe that $\delta\lambda=0$ is a maximum of $g(\delta\lambda)$, with $g(0)=0$, and a point where $f(0)=g(0)$. Is it possible that $f(\delta\lambda)\ge g(0)=f(0)$? Consider that there exists $\delta\lambda^*$ such that $f(\delta\lambda^*)< f(0)$. Then
\be
g(\delta\lambda^*)\le f(\delta\lambda^*)< f(0)=g(0),
\ee
leading to no contradiction again.

\section{Conclusion} 

I presented four fluctuation theorems related to the quasistatic work of a thermally isolated system. All of them were deduced based on Jarzynski's equality. Even though, none of them derived the strong inequality. Indeed, when Jarzynski's equality was required in these modifications, it did not offer a tight lower bound to the difference $\langle W\rangle-\langle W_{\rm qs}\rangle$. Also, assuming the negativeness of such a difference does not imply any contradiction with already known laws and those deduced in these fluctuation theorems. A piece of extra information must be provided to prove such postulate.

\section*{Conflict of interest statement}

There is no conflict of interest in the parts involved.

\section*{Data availability}

No data was generated in the present study.

There is no conflict of interest in the parts involved.

\bibliography{bibliography.bib}
\bibliographystyle{apsrev4-2}

\end{document}